\documentclass[english, aps, prb, twocolumn, noshowpacs, superscriptaddress, reprint]{revtex4-1}
\usepackage{color}
\usepackage{textcomp}
\usepackage{graphicx}
\usepackage{hyperref}

\begin{document}

\preprint{Tr/1.1-pre}

\title{Versatile AFM setup combined with micro-focused X-ray beam}

\author{T. Slobodskyy}
 \email{Taras.Slobodskyy@physik.uni-hamburg.de}
 \affiliation{Institut f\"ur Angewandte Physik und Zentrum f\"ur Mikrostrukturforschung,\\Jungiusstra\ss e 11, D-20355 Hamburg, Germany}
\author{A.V. Zozulya}
 \email{alexey.zozulya@desy.de}
 \affiliation{DESY Photon Science, Notkestra\ss e 85, 22607 Hamburg, Germany}
\author{R. Tholapi}
\author{L. Liefeith}
 \affiliation{Institut f\"ur Angewandte Physik und Zentrum f\"ur Mikrostrukturforschung,\\Jungiusstra\ss e 11, D-20355 Hamburg, Germany}
\author{M. Fester}
 \affiliation{DESY Photon Science, Notkestra\ss e 85, 22607 Hamburg, Germany}
\author{M. Sprung}
 \affiliation{DESY Photon Science, Notkestra\ss e 85, 22607 Hamburg, Germany}
\author{W. Hansen}
 \affiliation{Institut f\"ur Angewandte Physik und Zentrum f\"ur Mikrostrukturforschung,\\Jungiusstra\ss e 11, D-20355 Hamburg, Germany}

\date{\today}

\begin{abstract}

Micro-focused X-ray beams produced by third generation synchrotron sources offer new perspective of studying strains and processes at nanoscale. Atomic force microscope setup combined with a micro-focused synchrotron beam allows precise positioning and nanomanipulation of nanostructures under illumination. In this paper, we report on integration of a portable commercial atomic force microscope setup into a hard X-ray synchrotron beamline. Details of design, sample alignment procedure and performance of the setup are presented.

\end{abstract}



\maketitle

\section{Introduction}

Highly intense micro-focused X-ray beams are readily available at modern third generation synchrotron sources. Precise positioning of the synchrotron beam to the desired nanostructure can be a tedious process consuming precious synchrotron time. 

Multiple approaches have been developed for simultaneous analysis of nanostructures using Scanning Probe Microscopy (SPM) and X-ray methods. SPM techniques such as Atomic Force Microscopy (AFM) and scanning tunneling microscopy (STM) have already been combined with the micro-focused X-ray beams~\cite{scheler_probing_2009, pilet_nanostructure_2012, ishii_imaging_2008, fauquet_combining_2011, cummings_combining_2012, saito_development_2006, rose_new_2011, chiu_collecting_2008, wang_easy--implement_2013}. It has been shown that morphology and material sensitivity of the scanning methods combined with versatility of X-ray methods opens new opportunities for dedicated research of nanoworld.

The advantage of the AFM techniques~\cite{binnig_atomic_1986, giessibl_advances_2003} is that it provides the possibility to image, address and adjust infrastructures independent of the substrate conductivity, while STM works on conducting substrates only. Besides, similar to an STM tip the AFM-tip can be employed as a manipulation tool to allow precise and controlled modification of individual nanostructures and biological systems~\cite{scheler_probing_2009, schoenenberger_slow_1994, thalhammer_atomic_1997, fotiadis_imaging_2002}.

Implementation of an AFM setup at a synchrotron beamline environment implies  certain modifications of the AFM design. Therefore, most of the scanning probe microscopes implemented at synchrotron beamlines up to now are very specific and individually manufactured~\cite{scheler_probing_2009, fauquet_combining_2011}. 

Here we present design and operation of a commercial AFM setup working at ambient conditions installed at the coherence beamline P10 of PETRA~III synchrotron source, DESY Hamburg. The AFM setup provides contact, non-contact and dynamic modes of operation and shows good vibration stability while mounted on a goniometer table at the beamline. The resources needed for purchase and installation of the described AFM at a synchrotron beamline were comparable to installation of an average AFM setup elsewhere because we used a standard setup provided by the manufacturer. Special feature of the setup is the possibility to access the sample with X-ray beam in a range of incident angles up to 15 deg. Feasibility of measurements in conducting AFM modes such as Scanning Spreading Resistance Microscopy (SSRM)~\cite{avila_electrical_2010} and detection of photoelectrons in picoampere range has been demonstrated while the AFM was installed on the diffractometer.

\section{X-ray scattering geometry}

X-ray scattering configuration demands free propagation pathways for incoming and scattered X-rays. Interference of the sample environment with the beam can potentially hinder the data interpretation. Therefore, investigation of thin epitaxial films or objects located on a substrate introduces an additional experimental constrain of keeping the incoming and outgoing beams above the substrate horizon.

\begin{figure}
\includegraphics [width=8cm]{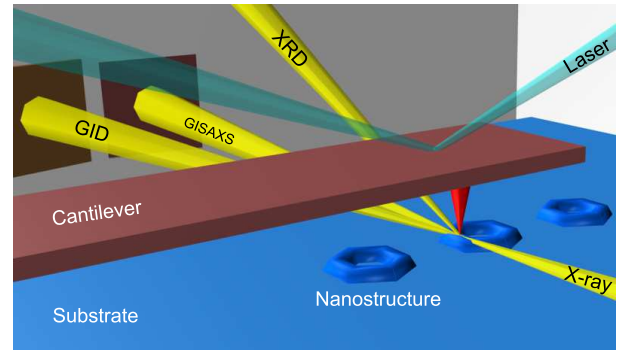}
\caption{\label{ScatteringGeometryAFM} AFM cantilever combined with X-ray scattering beam path.}
\end{figure}

\begin{figure*}
\includegraphics [width=17cm]{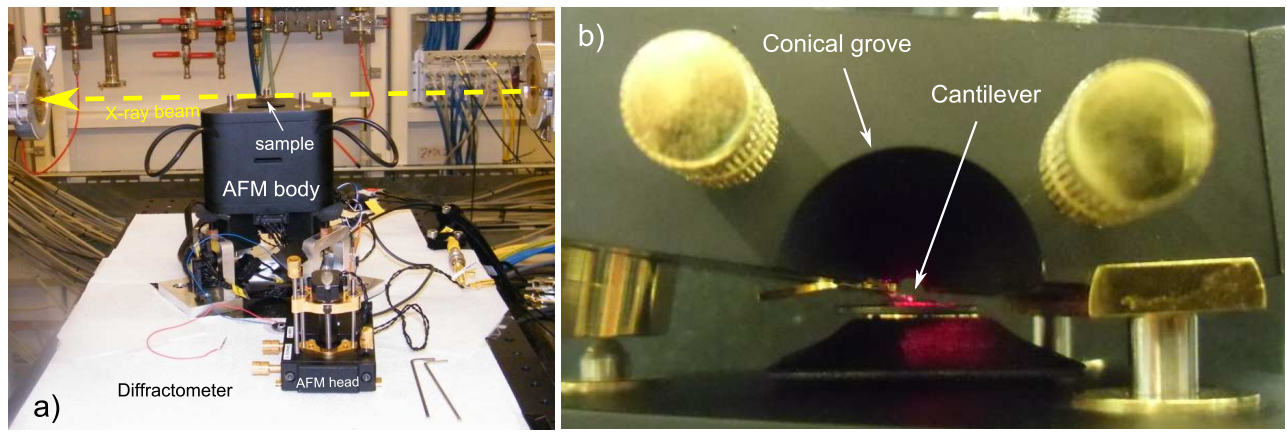}
\caption{\label{AFMbeamline} (a) Side view of the AFM body mounted on the diffractometer tower. (b) AFM tip contacting the sample as seen in the beam propagation direction.}
\end{figure*}

Let us now take a closer look on the geometrical constrains imposed by an AFM setup on the X-ray experimental conditions. Figure~\ref{ScatteringGeometryAFM} displays a schematic presentation of an AFM cantilever located on a substrate surface combined with incoming and outcoming X-ray beams. The cantilever deflection is measured with a laser beam which is reflected from the backside of the cantilever. The direction of incoming X-ray beam is perpendicular to the cantilever. The incoming beam satisfies conditions of grazing incidence X-ray scattering (GISAXS) as well as grazing incidence diffraction (GID) experiments. The corresponding exit beams are shown in Fig.~\ref{ScatteringGeometryAFM} propagating behind the cantilever. When an incoming beam is aligned at a Bragg angle to the surface it generates the diffracted outgoing beam (labeled XRD in Fig.~\ref{ScatteringGeometryAFM}). For this out-of-plane  diffraction conditions there might be further geometrical limitations due to the cantilever shape.


Since the particular AFM design has cantilever mounted at the head part of the AFM, the angular space available for X-ray beams to access the surface is limited by the cantilever and AFM head geometry (see Fig.~\ref{ScatteringGeometryAFM},~\ref{AFMbeamline}~b) ). Since the AFM tip cone is typically about 40{$^{\circ}$}, about 70{$^{\circ}$} out of the sample plane could be used for X-ray scattering experiments. But the tip is fixed on a cantilever which is much broader. Therefore, the maximal out-of-plane incidence and exit angles accessible in the direction perpendicular to the cantilever propagation direction are given by: {$ Q=\arctan {\frac{2 h}{w}} $}. Assuming a typical width of the cantilever  {$w = 30~\mu m$} and tip height {$h = 15~\mu m$} the maximal opening angle of 45{$^{\circ}$} is accessible for diffraction experiments. The out-of-plane angles along the cantilever propagation direction are even smaller. This direction is, however, unlikely to be used for out-of-plane diffraction measurements. Inclination of the tip relative to the surface may be used to provide an additional angular gain for incidence or exit angles when needed.

The angular space in-plane of the sample is virtually unconstrained. But we have to bear in mind that the AFM head should be rigidly connected to the AFM body and, therefore, the in-plane angular space will be constrained by the design of a particular AFM setup.

\section{AFM design and integration with the synchrotron beamline}

The task of simultaneous X-ray and AFM measurement of an individual nanostructure can be reduced to the two following tasks. First, one has to locate a particular nanostructure using an AFM tip. Second is to align the AFM tip with the sample into the X-ray beam. One way to solve this problem would be to fix the AFM on the diffractometer table and scan the sample with the X-ray beam by moving the whole AFM setup. At this configuration the AFM tip should be moved to keep its relative position to the beam constant. This is not a trivial problem and requires very precise control and synchronous operation of the AFM and the beamline equipment when the AFM setup is mounted on the beamline diffractometer. A more elegant solution is to move the sample relative to the AFM cantilever which is fixed relative to the beam position. The sample movement in this case is performed by the AFM sample translation stage. Since the sample movement does not modify the relative position of the AFM tip to the X-ray beam, it is sufficient to align the AFM tip relative to the beam by translating the complete AFM setup only once. Positioning of a defined nanostructure in the X-ray beam is done by positioning it in close proximity to the AFM tip.

In an AFM setup with fixed tip the scan is performed by moving the sample relative to the tip in all 3 directions. During a scan the AFM feedback mechanism keeps constant deflection of the cantilever. This means that the change in the sample morphology is compensated by a change in the vertical position of the sample during a scan. Constant deflection of the cantilever also means that the sample surface is at the same height relative to the X-ray beam. It is, however, important to keep in mind that the change in the AFM set point will influence the cantilever deflection and, therefore, the distance between the AFM tip and the X-ray beam.  The effect of set point change can be compensated by proper movement of the AFM setup using the diffractometer motors.


Let us discuss design and operation of the AFM setup used in our experiments. We are using an Anfatec Level AFM~\cite{ANFATEC_website}. The AFM is basically a standard setup with minor modifications which are aimed at extending angular space available for X-ray beam and increasing stability of the setup at inclinations. The AFM consists of two major parts: the body and the AFM head as shown in Fig.~\ref{AFMbeamline}. The sample is fixed on a piezo-driven stage on top of the body. The AFM head is placed on three motor driven shafts extending from the body. The cantilever is fixed on the bottom part of the head. This design permits flexible variation of distance between the sample and the tip. Since the three shafts can be moved independently, the inclination angle between the tip and the sample can be adjusted too. The AFM head can be easily removed leaving the sample open which is advantageous during the beam and sample alignment. Some other AFM manufacturers are also using similar approach leading to a flexible choice of possible suppliers.

During the sample alignment the AFM body is mounted on the diffractometer table while the head is removed (see Fig.~\ref{AFMbeamline}~a). The sample surface is aligned in the X-ray beam to produce the scattering or diffraction signal that will be further investigated. After the sample alignment, the AFM head is mounted on the AFM body and the tip is approached to the sample. At this point we need to ensure that the X-ray beam is illuminating the AFM tip apex.This is done by placing the sample surface parallel to the beam propagation direction, and scanning the sample surface by translating with the goniometer stage the whole AFM setup in the beam horizontally. The position at which the AFM tip is in the beam is found by the increase in the scattering signal originating from the tip edges. After this, the beam is placed at the end of the tip by vertical translation of the AFM setup. The end of the tip is found by approaching the sample horizon. The AFM together with the sample is tilted to the required incidence angle. Now, the beam is located just below the tip of the AFM. The diffractometer translations are fixed which ensures constant relative position between the beam and the tip. A nanostructure of interest is located by the AFM tip, thus bringing the nanostructure to the center of an X-ray beam.

A view of the AFM head during measurements is shown in Fig.~\ref{AFMbeamline}~b). The figure shows the AFM tip in contact with a sample as seen along the beam propagation direction. The backside of the AFM cantilever is illuminated by a red laser. The reflected laser beam is reflected from the backside of a cantilever, and the reflected beam is registered by a 4-segment photodiode. The segmented diode provides the 'top-minus-bottom' and 'left-minus-right' signals which are generated by deflections of a cantilever tip  in vertical and horizontal directions, correspondingly. Custom made conical grooves for the X-ray in the AFM head with the opening of 40{$^{\circ}$} can be seen in the figure. The head rests upon the three support shafts fixed by a ball and socket mechanism. The conical grooves enable access of X-rays to the sample under inclination angles of up to 20{$^{\circ}$}. The range of inclination angles is important for out-of-plane diffraction geometry which corresponds to the situation when the AFM is mounted on a 6-circle diffractometer\footnote{See supplemental material at [URL will be inserted by AIP] for details of the AFM setup operation on the six circle diffractometer of the P10 beamline as well as comparison of AFM images taken at different inclination angles.}.

As can be seen from Fig.~\ref{AFMbeamline}~b) the AFM support shafts constrain the accessible in-plane angular space. The shafts limit angular space available for in-plane diffraction and scattering experiments. We can see that the cantilever is tilted to project the tip closer to the sample plane as it is typical for an AFM deployment. The tilt of the cantilever, however, makes the angular range in the left direction shown in Fig.~\ref{AFMbeamline}~b) more suitable for diffraction in this direction. The single support shaft from the left side makes this direction more suitable for diffraction experiments.

The AFM setup is operated by a control unit and a separate high voltage power supply unit. An additional connection is made between the lock-in amplifier located inside of the control computer and the AFM body. During laboratory operation the AFM is levitated on three rubber stripes fixed to the AFM body at one end and attached to three support posts at the other end. The three support posts are rigidly fixed to a granite plate equipped with the necessary cable connections to the AFM controller, piezoamplifier and computer. This configuration provides good vibration isolation, the possibility of controlled sample humidity around the AFM by means of a jar bell and electrical isolation by means of a Faraday cage.

We have chosen to use a remote "VNC" control software to access the AFM computer~\cite{UltraVNC_website}. This configuration has proven to be the most reliable. The connection is made from the beamline control room and the operator is able to observe signals of the AFM in real time. At the same time it is beneficial to have the AFM control computer near the diffractometer. The AFM can be controlled locally from the experimental hutch during sample mounting and adjustment and remotely from the control room computer during X-ray experiments. Since no direct connection between the beamline control software and the AFM control computer was present, we synchronized computer times and correlated the X-ray data collected during the beamtime to AFM frames by the file creating times.

\section{Mounting and vibration stability}

Proper vibration isolation is crucial for an AFM setup. In laboratory conditions extreme caution is taken to suppress external vibrations in the AFM setup. As has been mentioned, the vibrations are normally suppressed by hanging the AFM body on three rubber stripes. This, however, is not possible during operation on a diffractometer table because a precise alignment of the sample in the beam is required. We have designed a special vibration damping support stage for placing the AFM on the diffractometer table for small angle scattering experiments. The mechanical part of the stage consists of three adjustable pivots which are rigidly fixed to the table. A 5 mm thick pad made of porous rubber is placed on top of each pivot to increase contact surface and enhance vibration damping. The AFM body is placed on the three pivots and is kept there by its weight as is shown in Fig.~\ref{AFMbeamline}~a).

\begin{figure*}
\includegraphics [width=17cm]{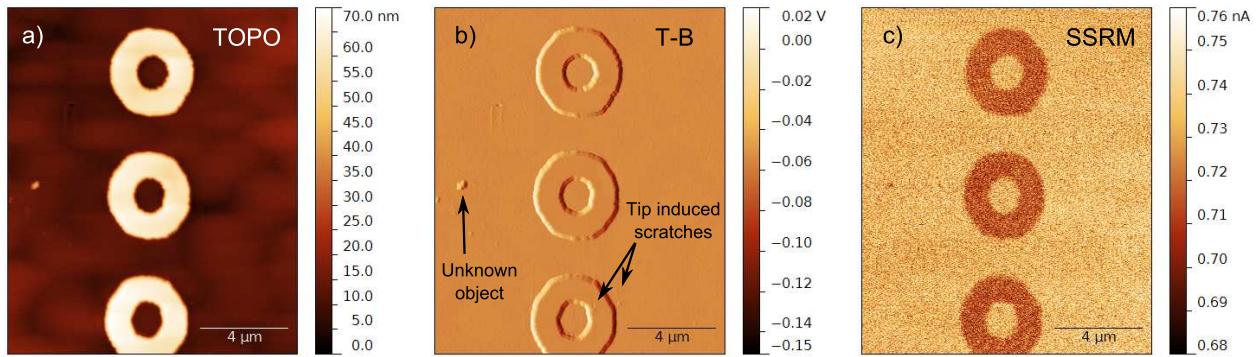}
\caption{\label{AFMscan} AFM image of the GaAs micro-ring sample obtained under conditions of the X-ray diffraction experiment. a) AFM topology. b: Vertical cantilever deflection (top-minus-bottom) signal. c) SSRM image}
\end{figure*}

The stage provides rigid connection between the diffractometer table and the AFM body. Careful tuning of the isolation pads material and thickness to increase the vibration damping was performed prior to transferring AFM to the beamline. After transferring the setup to the diffractometer table we found no significant increase of noise disturbances as compared to the performance of the AFM under laboratory conditions.

\section{Performance of the AFM}

Operation of the AFM on the diffractometer is demonstrated by  applying contact AFM and SSRM imaging to a test structure containing an MgO tunneling barrier at a metal/semiconductor interface. Such structures containing MgO barriers at Fe/GaAs interfaces are interesting for spintronics applications.

The sample was fabricated on a GaAs substrate using Molecular Beam Epitaxy (MBE). After oxide desorption a 300nm buffer layer was grown. A 500 nm GaAs conducting layer of with doping concentration of {$5 \times 10^{16} cm^{-3}$} serves as a spin channel. In the following 30~nm transition layer the carrier concentration is gradually increased to {$3 \times 10^{18} cm^{-3}$}. The top 15~nm are highly doped with silicon to provide sufficient amount of states available for tunneling.

The sample was then transferred by an Ultra High Vacuum (UHV) channel to a MgO deposition chamber. In the chamber a 0.5~nm MgO tunneling barrier was deposited by means of electron beam evaporation. The deposition was performed at room temperature under supply of molecular oxygen background.

For the last deposition step the sample was transferred to metal MBE chamber. In the UHV camber, the sample was annealed for 1~hour at 250~{$^{\circ}$}C. After cooling the sample a 5~nm thick iron ferromagnetic layer was deposited. The layer was followed by a 5~nm thick protection gold layer.

As a last fabrication step, the sample was taken out of the MBE system and processed using wet chemical etching into arrays of microring structures investigated here. For the structure preparation the top metal layers, the MgO tunneling barriers and 40~nm of GaAs were removed everywhere except below the ring mask. The arrays are aligned to allow access to individual microrings by the micro-focused X-ray beam.

The spin injection efficiency from the ferromagnetic top contact through the tunneling barrier is measured in the final structure. The highly doped top GaAs layer around the microring structure decreases the spin live time in the semiconductor and therefore was removed.

Let us first inspect the quality of the AFM image acquisition. Figure~\ref{AFMscan} shows an exemplary AFM scan of a microstructure array. The measurements were performed inside of the second experimental hutch of the P10 beamline. The images were taken in contact mode using all-diamond conductive ND-SSCRS cantilevers from Advanced Diamond Technologies. Presented signals were measured during a single forward scan (from left to right) of a double scan at the speed of 20~{$\mu m$} per second. 


Figure~\ref{AFMscan}~a) shows the topography of the sample surface. Post acquisition leveling of the data was performed. Three individual microrings can be observed on the AFM scan. The lithographically prepared rings are 52~nm high. The rings show outer radius of 1900~nm. The radius of the inner opening in the ring is about 700~nm. The structures are well resolved and the root mean square average R{$_{a}$} of the topography image is about 0.4~nm. Due to color scaling, however, not all morphological details can be observed clearly.

Fig.~\ref{AFMscan}~b) shows original vertical cantilever deflection (top-minus-bottom) signal received during the same scan. The signal represents the error function of the AFM feedback loop and, therefore, the observed contrast depends on the scan direction. Since we show only the forward scan direction the up slopes are brighter and the down slopes are darker. The values of the signal are in the range of values received while scanning in our well controlled laboratory conditions. A rich set of features can be observed in this vertical deflection scan. The most prominent surface feature apart from the rings is a 30~nm high object of unknown origin on the left. The rest of the features observed in Fig.~\ref{AFMscan}~b) are about 6~nm high. A tiny scratch can be seen on the upper left part of the bottom ring on careful examination. A similar scratch is visible near the ring on the GaAs surface. The scratches were made by applying force of 15~$\mu$N to the diamond cantilever during simultaneous X-ray and AFM investigations of the ring and demonstrate an example of nanoscale modification during the X-ray experiments. 

By using a conducting AFM tip and an electrical contact to the samples conducting channel we have collected SSRM images of the structures as shown in Fig.~\ref{AFMscan}~c). The SSRM image was collected simultaneously with the morphology scan. The image scale is 80~pA. A single parasitic frequency of 50~Hz was subtracted from the data by post processing. Interpretation of the image relies on the assumption that during an SSRM scan the resistance is measured at the point of the tip contact to the sample. The AFM image created by scanning the surface with the tip reflects lateral distribution of the sample conductivity~\cite{avila_electrical_2010}. The highly doped semiconductor surface shows a homogeneous conductivity as expected for a high quality MBE grown sample.

When the AFM tip is located over the top metal contact of the microstructure (see Fig.~\ref{AFMscan}~c) ) electrical contact is made between the tip and the top electrode. Since the resistivity of the top electrode is small as compared to the resistivity of the MgO tunneling barrier, the total resistance is dominated by the tunneling barrier located below the metal contact. This suppresses current variations during imaging the top metal contact.

\begin{figure*}
\includegraphics [width=17cm]{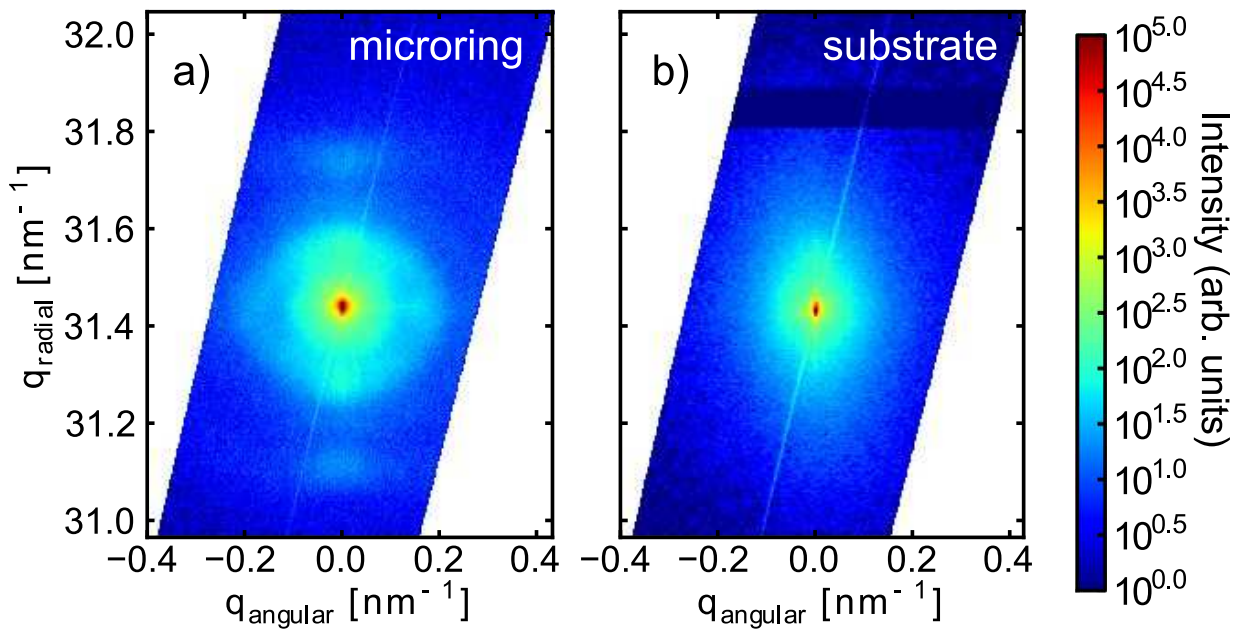}
\caption{\label{GidCompare} (a) Grazing incidence diffraction map of an individual micro-ring structure and (b) GID map measured from the plain substrate between two microrings.}
\end{figure*}

The contrast between the exposed semiconductor surface and the rings containing MgO tunneling barriers seen in Fig.~\ref{AFMscan}~c) is much smaller than the difference between the surface conductivities of the two. The reduction in the contrast is caused by presence of a highly resistive tunneling barrier which dominates electrical transport as described above. Taking into account the difference in the size of the active voltage drop area between the GaAs surface and the top of the rings the resistivity of about {$3 \times 10^{6} \Omega , \mu m^{2}$} corresponds to about 1~nm thick MgO tunneling barrier.

Vibrations induced by the stepper motors during the diffractometer movement between the measurement points were measured by the AFM tip which was kept in contact to the sample. The vibration amplitude was below 20~nm during the motor movement. Additionally, by comparing quality of the AFM scans before and after the diffractometer motor movement we observed no induced tip or sample damage. Therefore, it is safe to assume that reciprocal space maps of nanostructures under constant AFM tip induced strain can be collected using this setup without additional modifications.

In addition to the results reported above, we have designed an adapter unit equipped with motorized XYZ positioning stage for mounting the AFM setup on a 6-circle diffractometer located in the first experimental hutch of the P10 beamline. When the AFM setup is fixed inside of the 6-cicle diffractometer vibration stability similar to the conditions on the horizontal diffractometer is routinely reached. We also investigate the AFM setup operation on the six circle diffractometer of the P10 beamline as well as comparison of AFM images taken at different inclination angles~\cite{Note1}.

\section{GID measurements on individual spintronics structures}

In the second hutch of the coherence beamline P10 of PETRA III a two dimensional x-ray detector Pilatus 300K is located at the distance of 5~meters from the rotation center of the horizontal diffractometer. The diffraction arm can be rotated in horizontal plane up to 30{$^{\circ}$} 2-theta angle. Using the photon energy of 13~keV the (220) diffraction peak of GaAs with Bragg angle of 13.8{$^{\circ}$} can be measured in grazing incidence diffraction geometry. The X-ray beam was focused to a size of 1.5~x~3~{$\mu m^2$} (vertical~x~horizontal) using a set of parabolic refractive lenses as selected by transfocator optics \cite{zozulya_microfocusing_2012}. For the selected photon energy of 13~keV and focal distance of 1.6~m the lens set consisted of 8 lenses with  curvature radius of 50~{$\mu m$}. 

After the alignment procedures we focus on the individual microstrucure of the studied sample. For this, we located the particular microring shown in the bottom of the Fig.~\ref{AFMscan} and moved the AFM tip over the structure. The beam was propagating from the right to the left side of the figure. In the grazing incidence diffraction experiment reported here the footprint of the X-ray beam was extended in the beam propagation direction. Since the sample contained a linear sequence of microrings extended perpendicular to the beam propagation direction, we were able address individual structures without interfering with others.

Because the micro-focused X-ray beam was located below the AFM tip, we expect the small angle scattering signal from the microrings to be observed only when the tip is located above one of them. To verify our alignment we have performed several tests. By moving the tip between the rings the scattering signal originating from the rings was suppressed. And, during a two dimensional raster scan of the sample in the plane perpendicular to the beam propagation direction performed using the diffractometer table we were able to observe contrast in the integrated intensity map consistent with the shape of the AFM tip.

Prior to diffraction measurements, the sample was aligned for GISAXS conditions so that the beam is impinging the surface at twice the critical angle and the detector is positioned to detect directly scattered intensity. After GISAXS alignment the AFM setup was rotated horizontally to the Bragg angle position of the GaAs (220) GID reflection. To ensure that the AFM tip is located in the rotation center of the diffractometer the AFM tip was realigned using the small angle scattering signal as already described. Then the sample was tilted to the required incidence angle by observing specular reflection spot on the detector and the detector arm was rotated to the twice Bragg angle position.


Figure~\ref{GidCompare} presents GID reciprocal space maps taken while the AFM tip was located on a microring (a) and on an unpatterned substrate area between two microrings (b), respectively. The position of the sample in the beam was adjusted using AFM piezodrives while the AFM tip was stationary relative to the beam. The AFM tip feedback was used to monitor the sample movement. 

A characteristic circular distribution of diffuse scattering is visible around the (220) GaAs diffraction peak in Fig.~\ref{GidCompare}~a). The ring diameter is close to the reciprocal diameter of the microring and depends strongly on the exit detection angle. Similar dependence of the signal on the exit angle was observed during investigations of strain induced by self organized InAs quantum dots~\cite{schroth_investigation_2012}. This similarity indicates that the observed diffuse scattering feature could be originating from strain fields induced by the top MgO and metal layers on the top surface of the GaAs microring.

The Fig.~\ref{GidCompare}~b) shows a GID reciprocal space map measured while the center of the beam was located between two microrings. The diffraction signal is decaying smoothly and no pronounced features can be observed in the signal. By comparing the maps obtained from the ring and from the space between the rings we conclude that our setup allows reliable positioning of individual structures in the beam. Provided that the x-ray beam is well aligned relative to the AFM tip and the AFM piezo-driven stage is used for sample movement, sample positioning in the beam with sub-nanometer precision is feasible.

\section{Summary and conclusions}

In conclusion, we report on the design, operation and example application of an AFM setup integrated into the synchrotron beamline environment. The techniques of an AFM tip alignment relative to the micro-focused beam are analyzed. Mounting on a diffractometer, positioning and vibration isolation of the AFM setup discussed, the performance of our system is described and results of the setup operation are presented. By means of AFM alignment the sensitivity of GID measurements using micro-focused beam to individual spintronics structures has been demonstrated.

We have applied our AFM setup for aligning and imaging of an individual spintronics micro-structures using microfocused X-ray beam. Distinct signal originating from the microstructure was observed in a grazing incidence diffraction experiment. AFM contact mode was applied during the measurements. The contact mode is very suitable for in-situ nanomanipulation.

We hope that our descriptions will help other research groups to make proper decisions while designing experiments dealing with nanostructure investigation at ambient conditions.

\begin{acknowledgments}
The authors would like to thank PIER Ideenfonds for financial support via the project DIMAP and BMBF via the European project Era.Net.Rus "Spinbarrier".
\end{acknowledgments}

\end{document}